\documentclass[super,sort&compress]{elsarticle}
\usepackage{enumitem}
\usepackage{paralist}
\usepackage{todonotes}
\usepackage{hyperref}
\usepackage{rotating}
\usepackage{booktabs}
\usepackage{doi}
\usepackage[final,tracking=true,kerning=true,spacing=true,factor=1100,stretch=10,shrink=10]{microtype}
\usepackage[T1]{fontenc}

\usepackage{caption}
\usepackage{subcaption}

\makeatletter
\def\ps@pprintTitle{%
  \let\@oddhead\@empty
  \let\@evenhead\@empty
  \def\@evenfoot{\thepage\hfill}
}
\makeatother

\usepackage{fancyhdr}
\journal{Drug Discovery Today}

\setcitestyle{comma,super,open={},close={}} 
\makeatletter
\renewcommand\@biblabel[1]{#1.}
\makeatother

\begin{document}
% \linenumbers
\begin{frontmatter}

\title{Docking-based generative approaches in the search \\ for new drug candidates}
% \tnotetext[mytitlenote]{Fully documented templates are available in the elsarticle package on \href{http://www.ctan.org/tex-archive/macros/latex/contrib/elsarticle}{CTAN}.}

\author[ujaddress]{Tomasz Danel\corref{c1}}
\cortext[c1]{Corresponding author}
\ead{tomasz.danel@uj.edu.pl}
\author[ujaddress]{Jan {\L}\k{e}ski}
\author[panaddress]{Sabina Podlewska}
\author[ujaddress]{Igor T. Podolak}

\address[ujaddress]{Faculty of Mathematics and Computer Science, Jagiellonian University, Łojasiewicza 6, 30-348 Kraków, Poland.}
\address[panaddress]{Maj Institute of Pharmacology, Polish Academy of Sciences, Sm\k{e}tna 12, 31-343 Kraków, Poland}

\begin{abstract}
Despite the great popularity of virtual screening of existing compound libraries, the search for new potential drug candidates also takes advantage of generative protocols, where new compound suggestions are enumerated using various algorithms. To increase the activity potency of generative approaches, they have recently been coupled with molecular docking, a leading methodology of structure-based drug design. In this review, we summarize progress since docking-based generative models emerged. We propose a new taxonomy for these methods and discuss their importance for the field of computer-aided drug design. In addition, we discuss the most promising directions for further development of generative protocols coupled with docking.
\end{abstract}


\end{frontmatter}

\section{Introduction}

Drug discovery pipelines are nowadays connected (in the great majority of cases) with the application of various \textit{in silico} techniques, including machine learning (ML) methods, which can perform fast and effective analysis of huge amount of data\cite{Mitchell2014}.

Recently, ML was revolutionized by the rapid development of deep learning (DL) approaches, whose main characteristic is the ability to extract higher-level features from raw input data. Considering its huge potential, DL has also entered the field of computer-aided drug design (CADD). It has already been used for the development of quantitative structure-activity relationship (QSAR) models and in the virtual screening campaigns to improve the docking-based scoring of compound libraries\cite{Pereira2016}. In addition, DL methods were used to predict ligand-protein contacts\cite{Skwark2014} and for computational assessment of compounds physicochemical and ADMET properties\cite{Lusci2013}. 

DL bridged two fundamental approaches used in CADD: docking and generative methods. Structure-based approaches, with docking in particular, have already been the subject of several reviews and deep examinations~\cite{sulimov2021docking,pagadala2017software}. Also the generative approaches in general have already been quite well characterized in the context of their potential usage in the drug design process\cite{meyers2021denovomolecular}. Nevertheless, due to being a relatively new idea, the combination of generative models and docking has not yet been deeply explored despite being very important for the field of CADD. Its potential benefits to the process of design of new drug candidates include the possibility to explore novel chemical space on one hand (thanks to generative methods) and the instant assessment with reliable methods (thanks to docking algorithms).

\section{Background}

In this section, we provide a technical background on the tools used in generative modeling with molecular docking.

\begin{figure}[!tb]
\centering
\includegraphics[width=0.98\textwidth]{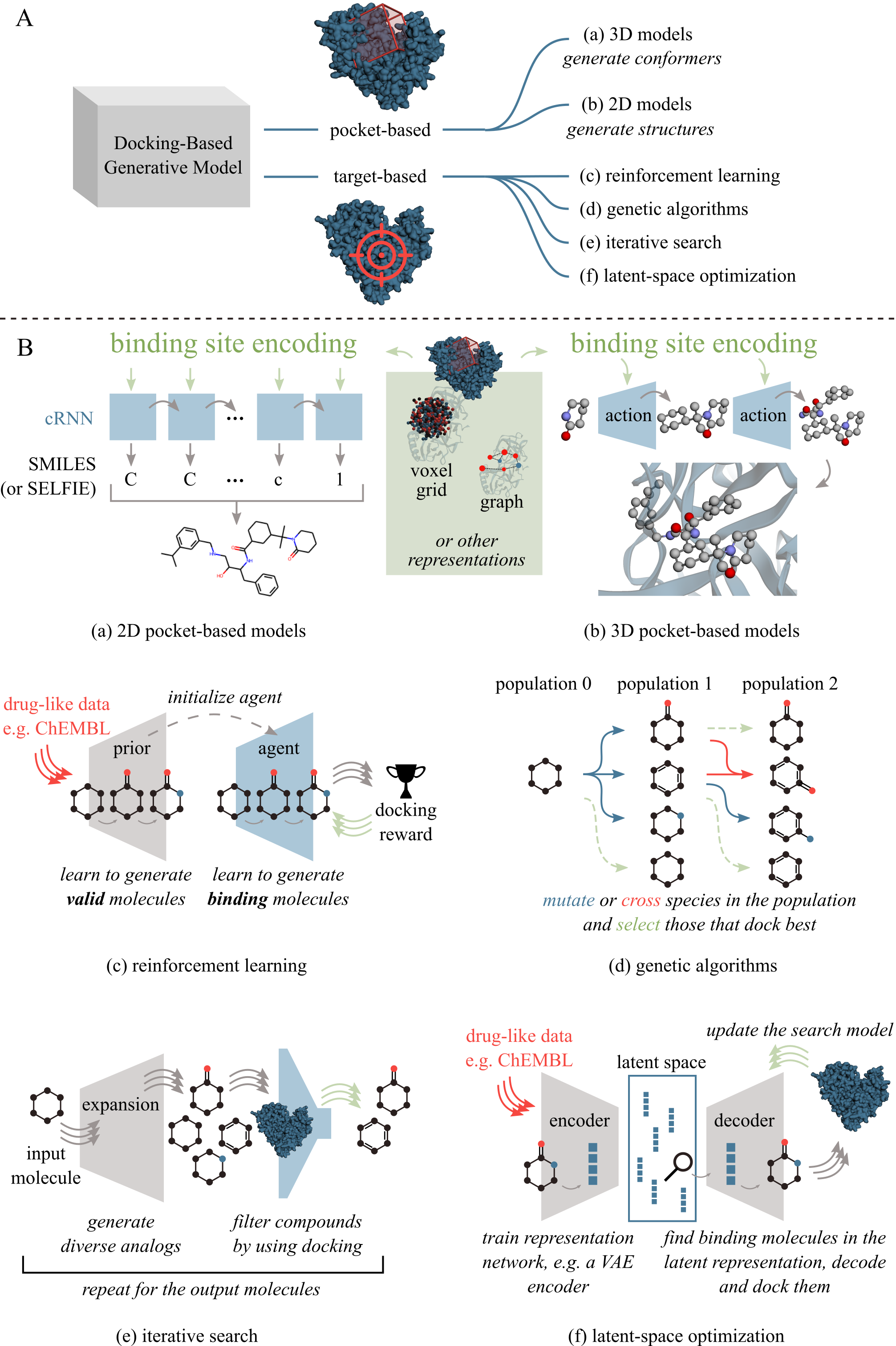}
\caption{The taxonomy of docking-based generative models proposed in the study.}
\label{fig:overview}
\end{figure}

\subsection{Generative models for molecules}

In molecular design, the term generative model describes a computer system that is able to yield new molecules, oftentimes possessing a set of pre-defined characteristics. Generative models in drug discovery are implemented to accelerate the design of novel therapeutic compounds\cite{sousa2021generative}.

Two molecular representations are prevalent in generative modeling. The linearized text representations, e.g. SMILES or SELFIES, can leverage the advancements in natural language processing, but they poorly describe geometrical relations between atoms. Molecular graphs can be of 2D and 3D type. The former ones capture the concept of atom neighborhood, whereas 3D graphs (conformers) exploit spatial relations thanks to the recent development of roto-translationally equivariant neural networks~\cite{han2022geometrically}.

Autoregressive models are one of the most widely used generative models in CADD~\cite{cheng2021molecular}. They build molecules out of atoms or fragments by selecting the most promising modifications in each step of the generation.
The next modification can be picked with the use of an oracle, e.g. predictive model or molecular docking. Otherwise, the model can be trained in a supervised manner by reconstructing the molecules in the input dataset.

Another class of generative models used in chemistry are latent-based models. In this approach, molecules are decoded from a continuous latent vector space, which is a vectorized molecular representation artificially constructed by data enumeration and model training. Autoencoders are a typical example of these models, including variational autoencoders~\cite{kingma2014auto} (VAE) and adversarial autoencoders~\cite{makhzani2015adversarial}. At first, they encode input molecules into a low-dimensional vector representation that follows one of the known probability distributions, and next they decode compounds from this representation to match the input molecules. Subsequently, if the decoder is an autoregressive network, it can be fine-tuned as described in the previous paragraph, or the latent space can be searched to identify potential ligands in a more manageable and structured data space. Generative adversarial networks~\cite{sousa2021generative} (GANs) are another example of latent-based models, but instead of using encoders, they use disciminator networks to evaluate compounds decoded from the sampled latent vectors.

\subsection{Molecular docking software used in the docking-based generative approaches}

Generative models started to leverage molecular docking to propose new drug candidates. Most widely used docking software is AutoDock Vina and its derivatives. It is an open-source program, under a permissive Apache license, widely-used with docking-based generative models. 
Other include smina\cite{koes2013lessons} -- a~branch of AutoDock Vina with an improved scoring function,  QuickVina\cite{alhossary2015fast} -- uses heuristics to accelerate docking, Glide\cite{friesner2004glide} -- an extensive and accurate tool, gnina\cite{mcnutt2021gnina} -- employing neural networks (NN) as a scoring function. NNs are sometimes trained on docking scores and used instead of docking to accelerate the process~\cite{choi2021v-dock,cieplinski2020we}. 

\subsection{Model evaluation metrics}

The fundamental metrics used to evaluate generative models are validity, uniqueness, and novelty of the generated compounds~\cite{brown2019guacamol}. Additionally, measures of drug likeness such as quantitative estimate of drug-likeness (QED) or synthetic accessibility (SA) (e.g. estimated by the molecule.one tool~\cite{steinmann2021using}) are used to assess the quality of these molecules. In most drug discovery projects, it is also important to have a diverse set of drug candidates, which can be measured by internal diversity or sphere exclusion diversity (SEDiv)~\cite{thomas2021comparison}. Finally, the most important metrics for the targeted generative models are docking-based measures~\cite{cieplinski2020we}.

For 3D generative models, additional measures verify the correct conformation of the generated molecules, e.g. 3D maximum mean discrepancy value or RMSD between the generated and reference conformations~\cite{li2021structure}. For some models, the shape and pharmacophoric constraints are checked using metrics like shape and color similarity score proposed by Imrie et al.~\cite{imrie2021deep3dpharmacophoric}.

\section{Docking-based generative models in drug design}

A docking-based generative model creates a number of drug-like molecules while enhancing their binding affinity by utilizing a computational docking model. Docking scores may be used directly, e.g. as a component of an optimized reward value, or indirectly as a filtering method, in which from a number of generated structures only those with a favorable docking score are retained. 

We partitioned the docking-based generative models into two categories: the pocket- and target-based models (see Figure~\ref{fig:overview}). The former construct a description of the binding pocket and create compounds that best fit the described binding site, employing either a 2D representation or building 3D molecular graphs directly inside the pocket. The target-based models are trained specifically for the selected drug target. They can be guided by reinforcement learning, genetic algorithms, or different iterative methods. Other algorithms explore the latent representation of molecules to identify potential binders of the given target in pre-trained generative models.

\subsection{Pocket-based models}

Pocket-based models use the shape and physicochemical properties of binding sites, either by encoding them in the model or by using docking scoring functions to assess the generated conformers (see Figure~\ref{fig:overview}a,b). Binding pockets can be represented, e.g., as 3D molecular graphs~\cite{luo20213d,li2021structure,peng2022pocket2mol,xia2019graphbased,han2022geometrically} or voxel grids~\cite{skalic2019target,ragoza2022generating}, and processed by graph (GNNs) or convolutional neural networks (CNNs), respectively. This way, the target protein can be replaced, in some cases, without model retraining, and the application of the model can be directly transferred to another target~\cite{luo20213d,ragoza2022generating,peng2022pocket2mol}.

\subsubsection*{SMILES models conditioned on binding pockets}
Xu et al.~\cite{xu2021denovomoleculedesign} encode binding pockets with eigenvalues of the Coulomb matrix of coarse-grained atoms and generate compounds with a conditional RNN. In an extension, Zhang et al.~\cite{zhang2022denovomoleculedesign} apply interaction fingerprints of the ligand-protein complexes obtained in docking, using an LSTM network to generate compounds. Both produce SMILES strings.

Zheng et al.\cite{zheng2021deep} implemented a transformer architecture for scaffold hopping using protein encoding as a context for the optimization. Proteins were encoded from the strings of amino acids using TAPE, which is also a transformer architecture.

Skalic et al.~\cite{skalic2019shape,skalic2019target} employed a semi--3D approach to generate compounds, where they encode and decode only the shape of a molecule, using a shape captioning network to decode SMILES strings. 3D CNNs encode the shape and pharmacophoric features of the molecule~\cite{skalic2019shape}.
Proteins can be encoded in a~similar way to ensure that ligands are matching the given protein description~\cite{skalic2019target}. BicycleGAN is employed to generate diverse ligands for a single input protein.

\subsubsection*{3D generative models}

The first report of a deep generative model producing three-dimensional compound structures conditioned on the receptor binding site was authored by Ragoza, Masuda and Koes\cite{ragoza2022generating}. As a starting point, the 3D bound molecular structure (including the target binding pocket) is encoded to a~latent space using atomic density grids transformed by CNNs. Atoms are encoded as continuous, Gaussian-like densities in a three-dimensional grid with separate channels for each atom type. Conditional variational autoencoder (CVAE) with the GAN loss function is used for training (the network was trained on the CrossDocked2020 dataset). Ligand and receptor density grids form a CVAE input and ligand density grid is returned as an output. In order to obtain valid molecules from the generated grids, an algorithm combining beam search and gradient descent is used to return a set of atom types and coordinates with the best fitting to a~given atomic density grid, which are finally combined into a valid molecule by the respective bonds' assignment. 

Luo~et~al.\cite{luo20213d} developed a 3D generative model which predicts the probability of atom presence in the binding site region. The atoms of the protein and ligand are encoded as a graph of atoms connected by the k-NN algorithm, and rotationally-invariant GNNs are used to create an atom probability density in the 3D space. Next, the autoregressive sampling algorithm is used for enumerating molecules from the model using the estimated density.

Li et al.~\cite{li2021structure}, instead of encoding binding pockets, directly incorporate docking to score the generated molecules. They develop an autoregressive method that consists of two networks: a state encoder and a policy network. The state encoder is a GNN that creates a representation of a partially built molecule, and the policy network decides what should be added to the molecule. MCTS with docking scores as the scoring function is used to sample molecules.

Peng et~al. developed Pocket2Mol\cite{peng2022pocket2mol}, an efficient system based on an E(3)-equivariant generative network, which combines a GNN capturing chemical and geometrical constraints of the target binding pocket and a sampling algorithm leading to the generation of novel ligand candidates conditioned on the 3D pocket. Pocket2Mol adopts the autoregressive strategy to learn a probability distribution of particular atom or bond type for a given space fragment inside a pocket on the basis of the already existing atoms (part of the drug is randomly masked, and the model is trained to predict the remaining fragment). 

\subsection{Target-based models}

In this section we describe different approaches to docking-based compound generation, used by target- and, to some extent, pocket-based models (see Figure~\ref{fig:overview}c-f).

\subsubsection*{Genetic algorithms GA}
GAs are fit for use with a docking score as the fitness function and have the advantage of not having to be trained.  Well known is GANDI, a fragment-based model\cite{dey2008fragmentbased}, in which pre-docked fragments are encoded by genetic algorithm and tabu search is used to find optimal linkers for the formation of final compounds (the fitness is computed as a combination of local binding energy and similarity to pocket).
Steinman and Jensen\cite{steinmann2021using} extended the graph based GB-GA model\cite{jensen2019gbga} to use the roulette selection and the fitness function to be a combination of docking score from Glide together with an index of synthesizability. Results found prove to be better than those of high throughput virtual screening.

AutoGrow4~\cite{spiegel2020autogrow4} is a large open-source GA-based package, fit for both \textit{de-novo} and lead optimization tasks, each starting with a different population. AutoGrow4 implements all generation operators (elitism, mutation, and crossover) using Ranking, Roulette and Tournament selections not to get trapped in local solutions. It uses a docking score (Vina and QVina \cite{alhossary2015fast} dockers are applied) together with diversity as its fitness score. Fu et~al.\cite{fu2022reinforced} notice that GAs have high variance across runs, due to the randomness in crossover and mutation.
Therefore, based on the AutoGrow4 environment, they propose to use separate equivariant NNs\cite{satorras2021equivariant} to first choose parents for crossover and mutation, and next the second parent for crossover or a reaction from the SMARTS set for mutation. The networks are trained using RL with the REINVENT scheme\cite{olivecrona2017molecular} using docking scores as a reward. This approach proves to give superior results, slightly better than AutoGrow4, REINVENT, and GB-GA.

The JANUS model\cite{nigam2022parallel} is interesting in that it uses two populations: one for chemical space exploration controlled by DNNs with crossover and mutation, and one for exploitation controlled only by mutation. The DNN is trained for each population using molecules with known docking values. The populations share the best members. JANUS uses SELFIES representations and the STONED model (see section on iterative models) for the definition of both of these operators, speeding up the generation of new members considerably.  Using the docking value as a function of fitness, the model achieves significantly low docking scores.
Importantly, the authors note that such a GA model must also have built into the fitness function the values responsible for the synthesizability and stability of the generated molecules. The member generation operators should also take care of these characteristics.

Designing the V-dock model, Choi et~al.\cite{choi2021v-dock} extended the MolFinder system\cite{kwon2021molfinder} by adding a NN trained to evaluate docking scores. The entire system uses SMILES, along with the CSA algorithm\cite{lee97newoptimization}, a combination of GA and simulated annealing. All of these elements greatly accelerate the model.

Popular with the medicinal chemists is another GA tool, LigBuilder\cite{yuan2020ligbuilder3}. LigBuilder~V3 offers a polypharmacological approach to ligand enumeration. It is able to generate compounds with reference to their activity towards multiple targets using the Chemical Space Exploring Algorithm (CSEA). CSEA constructs potential ligands starting from the placement of an sp3 carbon atom in a~randomly selected point of a binding site. Then, newly formed molecules are split into fragments, and those fragments which possess the highest potential of desired biological activity are used as a starting point for the subsequent growing operations.

\subsubsection*{Reinforcement learning models RL}\label{sec:reinvent}

Reinforcement learning (RL) has been used for more focused exploration of the chemical space in search of binding molecules. Docking scores can be used as rewards for the generated molecules to guide the generative process. Jeon and Kim~\cite{jeon2020autonomous} proposed an algorithm that builds a molecule by adding atoms and bonds sequentially. The intermediate steps are assessed in terms of SA and QED, and only the final compound is docked to evaluate its binding potential. Yang et al.~\cite{yang2021hit}, on the other hand, proposed to build compounds out of chemically reasonable fragments, and docking scores are computed for each generation step.

Several target-based generative methods \cite{thomas2021comparison} are based on the REINVENT model proposed by Olivecrona et al. \cite{olivecrona2017molecular} which uses RL for output optimization. First, a generative RNN model (the Prior) is trained on a subset of ChEMBL to generate valid SMILES strings. Then an Agent, a copy of the Prior, is trained with RL to modify the Prior's proposition towards some specific goal, e.g.~improved docking scores.
REINVENT~2.0\cite{blaschke2020reinvent} added diversity filters memorizing structures with similar scaffolds to drive the model towards higher diversity. Transfer learning added in the Prior training directed the creativity towards a~given molecule subset. LibINVENT\cite{fialkova2022libinvent} added moieties driving the RL module towards best docking, diversity or synthesizability.

Link-INVENT\cite{guo2022linkinvent} was designed for the fragment-based drug discovery problem: on the basis of a batch of fragments, the encoder-decoder proposes whole linked molecules to be optimized with RL towards best docking (Glide used). In the case of complex multi-parameter optimization (MPO), curriculum learning (CL) might be favorable\cite{guo2022improving}: divide optimization into several production objectives, e.g.~molecules with a given target scaffold, that are drug-like, that optimize some given feature, realized by the Agent sequentially.

Notably, DockStream~\cite{guo2021dockstream} is a docking platform providing protein and ligand preparation tools and multiple docking backends that can be combined with REINVENT for structure-based drug design (SBDD).

\begin{table}[htb]
\centering
\scriptsize
\caption{Docking-based generative models, including names, references, source code links, and licenses.}
\begin{tabular}{@{}l@{\;}l@{\;}l@{}}
\toprule
Name & URL & Licence\\
\midrule
3D SBDD model  \cite{luo20213d} & \href{https://github.com/luost26/3D-Generative-SBDD}{github.com/luost26/3D-Generative-SBDD} & MIT \\
Pocket2Mol  \cite{peng2022pocket2mol} & \href{https://github.com/pengxingang/Pocket2Mol}{github.com/pengxingang/Pocket2Mol} & MIT \\
LiGAN  \cite{ragoza2022generating} & \href{https://github.com/mattragoza/liGAN}{github.com/mattragoza/liGAN} & GNU GPL 2.0 \\
DeepHop  \cite{zheng2021deep} & \href{https://github.com/prokia/deepHops}{github.com/prokia/deepHops} & MIT \\
LigDream  \cite{skalic2019shape} & \href{https://github.com/compsciencelab/ligdream}{github.com/compsciencelab/ligdream} & AGPL-3.0 \\
GB-GA  \cite{jensen2019gbga,steinmann2021using} & \href{https://github.com/jensengroup/GB_GA}{github.com/jensengroup/GB\_GA} & MIT \\
AutoGrow4  \cite{spiegel2020autogrow4} & \href{http://durrantlab.com/autogrow4}{durrantlab.com/autogrow4} & Apache 2.0\\
JANUS  \cite{nigam2022parallel} & \href{https://github.com/asparu-guzik-group/JANUS}{github.com/asparu-guzik-group/JANUS} & Apache 2.0\\
V-dock \cite{choi2021v-dock} & \href{https://github.com/knu-chem-lcbc/V-dock}{github.com/knu-chem-lcbc/V-dock} & Open source \\
MolFinder \cite{kwon2021molfinder} & \href{https://github.com/duaibeom/MolFinder}{github.com/duaibeom/MolFinder} & BSD 3-Clause \\
LigBuilder V3  \cite{yuan2020ligbuilder3} & \href{http://www.pkumdl.cn/ligbuilder3/}{www.pkumdl.cn/ligbuilder3/} & Academic \\
MORLD  \cite{jeon2020autonomous} & \href{https://github.com/wsjeon92/morld}{github.com/wsjeon92/morld} & Apache 2.0 \\
FREED  \cite{yang2021hit} & \href{https://github.com/AITRICS/FREED}{github.com/AITRICS/FREED} & Apache 2.0 \\
MolScore~\cite{thomas2021comparison} & \href{https://github.com/MorganCThomas/MolScore}{github.com/MorganCThomas/MolScore} & MIT\\
REINVENT  \cite{olivecrona2017molecular} & \href{https://github.com/MarcusOlivecrona/REINVENT}{github.com/MarcusOlivecrona/REINVENT} & MIT\\
REINVENT 3.2  \cite{blaschke2020reinvent,guo2022linkinvent,guo2022improving} & \href{https://github.com/MolecularAI/Reinvent}{github.com/MolecularAI/Reinvent} & Apache 2.0\\
LibINVENT  \cite{fialkova2022libinvent} & \href{https://github.com/MolecularAI/Lib-INVENT}{github.com/MolecularAI/Lib-INVENT} & Apache 2.0\\
Link-INVENT  \cite{guo2022linkinvent} & \href{https://github.com/MolecularAI/Reinvent}{github.com/MolecularAI/Reinvent} & Apache 2.0\\
DockStream  \cite{guo2021dockstream} & \href{https://github.com/MolecularAI/DockStream}{github.com/MolecularAI/DockStream} & Apache 2.0\\
STONED  \cite{nigam2021beyond} & \href{https://github.com/aspuru-guzik-group/stoned-selfies}{github.com/aspuru-guzik-group/stoned-selfies} & Apache 2.0\\
JTVAE sample-and-dock  \cite{xu2021navigating} & \href{https://github.com/atfrank/SampleDock}{github.com/atfrank/SampleDock} & GNU GPL 3.0 \\
OptiMol  \cite{boitreaud2020optimol} & \href{https://github.com/jacquesboitreaud/OptiMol}{github.com/jacquesboitreaud/OptiMol} &  Open source\\
Gradient Latent Search  \cite{cieplinski2020we} & \href{https://github.com/cieplinski-tobiasz/smina-docking-benchmark}{github.com/cieplinski-tobiasz/smina-docking-benchmark} & MIT \\
\bottomrule
\end{tabular}
\label{tab:links}
\end{table}

\subsection{Iterative models with docking-based evaluation}\label{sec:iterative}
Some models have abilities to produce new molecules, but do not have means, other than heuristic, to optimize them towards better docking. We call them iterative here because the generated molecules are evaluated towards some target and filtered accordingly.

Ghanokta et~al.\cite{ghanokta2020combining} propose a model where PathFinder\cite{konze2019pathfinder} performs a retrosynthetic analysis starting from some molecule and enumerating all possible reactions, while keeping the core fixed. After filtering (SMARTS, physicochemical, duplicate removal) and docking against a given target, a subset is used to train an ML model based on the free energy perturbation (FEP) experiments, with input poses evaluated via Glide. This FEP prediction model is then used to filter numerous generated molecules and select the top ones. This path might be used to train the REINVENT Prior model, giving high correct core and reaction-group molecules, speeding up the quest for the best one. 

One possible improvement is to use the SELFIES representation\cite{nigam2021beyond}. The STONED model combines a sequence of point-wise modifications in a SELFIES to generate a large number of all valid new molecules (unlike when SMILES is used). This enables rapid generation of molecule subspaces by "superimposition" of point modifications and extend quicker from the original one. The authors also show how to produce unambiguous paths between two given molecules. As the system is discrete, it is not possible to directly optimize some parameters, e.g. the docking score, other than by using some heuristics.

Another model is based on Monte-Carlo Tree Search (MCTS)\cite{srinivasan2021artificial}. MCTS builds a search tree by adding a single level to a SMILES string in each selection-expansion stage. A high number of randomly generated complete strings are then evaluated with the Vina docker in the simulation stage. All nodes have their current score updated based on these evaluations (the backpropagation). The symbols which resulted in better scores are then chosen more frequently in later selection stages. Authors have identified a large fraction of molecules with better scores for a given target, than those from the FDA list. 

A similar in operation sample-and-dock approach\cite{xu2021navigating} uses a pretrained JTVAE model to map some molecule (the authors used benzene) to the latent space and then sample from its neighborhood. Sampled vectors are decoded, re-coded as graphs and docked (RDKit and rDock were used) against the target. The best molecules are mapped again in a similar loop until no better molecules can be found. It is possible to apply transfer learning after pre-training using an additional subset of specific active molecules. This needs to be done with care, not to hinder JTVAE generativity.

\subsection{Latent-space optimization}

Conversely to most of the methods described in the previous sections, some models use the latent space of already trained generators to discover new binders without changing the generative process itself. Any generative model that uses a continuous latent representation to generate molecules, e.g. an autoencoder, can be used as a prior generative model and can be trained on a big dataset of drug-like molecules, e.g. ChEMBL or ZINC. Next, a latent-space optimization algorithm is used to effectively sample latent variables that map to well-binding ligands.

An example of this approach is OptiMol\cite{boitreaud2020optimol}, an algorithm that uses a VAE as the prior generative model. First, the VAE is trained to encode molecular graphs and decode SELFIES representation. Next, two latent-space exploration schemes are used: Bayesian Optimization (BO) and Conditioning by Adaptive Sampling (CbAS). In both schemes, molecular docking is used to evaluate compounds decoded from the latent space and change the sampling strategy. BO uses Gaussian processes to find the most promising regions of the latent space, evaluate compounds sampled from these regions, and adjust the Gaussian process. %The authors argue that BO is intractable in the vast chemical space. 
CbAS, on the other hand, shifts the initial distribution of the prior generative model to maximize the docking-based objective function, which makes the search more effective.

Another approach to the latent-space optimization is presented by Ciepliński et~al.\cite{cieplinski2020we} who trained two VAE models, SMILES VAE and Grammar VAE, as prior generators and use a predictive model trained to map latent vectors to docking scores as a surrogate function. This way, new binders can be found in the latent space by gradient ascent maximizing model predictions starting from a random initial point in the latent space.

\section{Other models and future outlook}
There is still a number of models that do not address the problem of docking score optimization directly, but they are able to generate novel molecules and optimize their features, which may include docking objectives. For example, MOLUCINATE\cite{arcidiacono2021molucinate} is a VAE model that is able to produce a 3D molecule grid representation using graph spatial convolutions. The generated 3D positions could be paired with docking to provide active conformations. A deep 3D linker model\cite{imrie2020deep3dlinker,imrie2021deep3dpharmacophoric} is designed to create linkers between two small molecular graphs by adding bonds in a breadth-first manner. 
This model is evaluated using docking scores, which could be integrated in the training procedure. Yet another example is a conditional 3D generative model by Gebauer et al.\cite{gebauer2022inverse} For a given atom, the model generates a new one based on a factorized conditional probability. As conditions, the model uses isotropic polarizability, molecular fingerprints, and atomic composition, all embedded with an MLP network. Docking scores of the conformers could be added as another condition. A following idea of generating synthesizable molecules was suggested\cite{moret2020generative,grisoni2021combininggenerative}: first generate SMILES representation, then predict all reactions needed and filter molecules with those available, and then synthesize on a microfluidics platform. This model can easily be extended to optimize docking scores.

Other possible extensions of the current methods are combining generative
models with deep learning solutions for affinity prediction or the prediction of
docking poses. This would open up a path to fully-differentiable optimization of
chemical structures.

\section{Conclusions}
Generative ML methods constitute an important component of the set of the computer-aided drug design tools. They enable exploration of novel chemical space (in comparison to the commercially available compound libraries, which usually undergo virtual screening), and therefore, their popularity in the protocols for searching for new drugs is constantly increasing. In order to increase the effectiveness of generative methods to propose compounds which are active towards particular target, their combination with molecular docking has recently been intensively explored. In the review, we summarized the docking-based generative methods and proposed their taxonomy. Moreover, we propose the future directions in which the docking-based generative models can further be developed.

\section*{Acknowledgments}

This work was supported by the National Science Centre (Poland) grant no.~2020/37/N/ST6/02728.

\bibliographystyle{ama}
\bibliography{mybibfile}

\end{document}